\documentclass[9pt,twocolumn,twoside]{optica}
\setboolean{shortarticle}{true}
\setboolean{minireview}{false}

\title{Enhanced Room Temperature Infrared LEDs using Monolithically Integrated Plasmonic Materials}

\author[1]{Andrew Briggs}
\author[1]{Leland Nordin}
\author[1]{Aaron J. Muhowski}
\author[2]{Evan Simmons}
\author[3]{Pankul Dhingra}
\author[3]{Minjoo L. Lee}
\author[2]{Viktor A. Podolskiy}
\author[2]{Daniel Wasserman}
\author[1,*]{Seth R. Bank}

\affil[1]{Microelectronics Research Center and ECE Dept., The University of Texas at Austin, 10100 Burnet Rd. Bldg. 160, Austin, TX 78758, USA}
\affil[2]{Department of Physics and Applied Physics, UMass Lowell, One University Ave, Lowell, Massachusetts 01854, USA}
\affil[3]{Holonyak Micro and Nanotechnology Lab and ECE Dept., University of Illinois at Urbana-Champaign, Urbana, Illinois 61801, United States}

\affil[*]{Corresponding author: sbank@utexas.edu}

\dates{Compiled \today}

\doi{}

\begin{abstract}
Remarkable systems have been reported recently using the polylithic integration of semiconductor optoelectronic devices and plasmonic materials exhibiting epsilon-near-zero (ENZ) and negative permittivity. In traditional noble metals, the ENZ and plasmonic response is achieved near the metal plasma frequency, limiting plasmonic optoelectronic device design flexibility. Here, we leverage an all-epitaxial approach to monolithically and seamlessly integrate designer plasmonic materials into a quantum dot (QD) light emitting diode (LED), leading to a ~5.6 ${\times}$ enhancement over an otherwise identical non-plasmonic control sample. The device presented exhibits optical powers comparable, and temperature performance far superior, to commercially-available devices.
\end{abstract}

\begin{document}

\maketitle

Polylithic integration of semiconductors with plasmonic materials exhibiting epsilon-near-zero (ENZ) and negative permittivity response offers exciting prospects for enhancing light-matter interactions to increase optoelectronic device performance and realize new functionality  \cite{Vuckovic:03,Lau:09,Kim:11,Eggleston:15,Tsakmakidis:16}. Unfortunately, with traditional noble metals, ENZ and plasmonic response are achieved only at limited spectral positions near the metal plasma frequency, typically in the UV/visible wavelengths  \cite{Vuckovic:00,Fleury:13,Maas:13,Vassant:12}. This limits plasmonic optoelectronic device design and spectral flexibility, often leading to weaker performance improvements than predicted  \cite{Khurgin:12}. In the mid-infrared (mid-IR), however, the picture is very different, and the opportunity exists for designer plasmonic and quantum-engineered optoelectronic materials in the same material systems.

Molecular beam epitaxy (MBE) allows for the engineering of so-called “designer metals,” relatively low-loss plasmonic materials with plasma wavelengths that span the mid-infrared by control of doping concentration during growth \cite{Law:14}. The epitaxial materials typically employed as designer metals are narrow bandgap materials, such as InAs or InAsSb, whose small effective masses allow for plasma wavelengths as short as ${\lambda_p}$   ${\sim}$ 5 ${\mu}$m, where ${\lambda_p = 2 \pi c \sqrt{m^* \epsilon_o \epsilon_b / e^2 n_e}}$ [corresponding to the epsilon-near-zero (ENZ) wavelength where the doped semiconductor transitions from a lossy “dielectric” to a plasmonic “metal”], with ${m^*}$ being the electron effective mass, ${\epsilon_b}$ the background dielectric permittivity of the semiconductor, and ${n_e}$ the electron doping concentration. Additionally, and perhaps even more tantalizing, the InAs(Sb) material system is home to a variety of mid-IR emitting optoelectronic materials whose emission overlaps spectrally with the plasmonic and ENZ behavior of heavily doped InAs(Sb). Quantum cascade lasers, interband cascade lasers, superlattice-based emitters and nanostructured quantum dot materials, provide opportunities to monolithically integrate – with the atomic precision enabled by epitaxial growth – quantum-engineered emitters with these designer plasmonic materials \cite{Yang:97,Teissier:04,Zhang:95,Das:05,Yu:17}. In fact, a variety of quantum cascade lasers have leveraged noble metal waveguides to confine optical modes in lasers, and have even employed doped semiconductors to compensate unwanted positive group velocity dispersion \cite{Sirtori:98,Sirtori:99,Bidaux:17}. InGaSb type-II quantum dots grown in an InAs matrix are a particularly intriguing option due to their excellent temperature performance, a result of decreased Auger recombination rates, and increased radiative transition rates, stemming from their low dimensionality \cite{Levesque:11,Zabel:15,Briggs:20}. The wavelength flexibility of both the designer metals and In(Ga)Sb quantum dots allow for an epitaxially-integrated architecture for cavity enhancement, providing a highly-engineerable mechanism for improved output and efficiency across the mid-IR. Mid-IR LEDs have long lagged behind lasers in the same wavelength range. This is partially because the most efficient mid-IR lasers are weak sub-threshold emitters, making mid-IR LEDs (that utilize mid-IR laser active regions) extremely inefficient \cite{Faist:07}.

Transitioning a cavity-enhanced emitter architecture from optical pumping to an electrically-driven device has traditionally proven problematic.  In the case of an optoelectronic material (semiconductor emitter) coupled to a plasmonic structure, the electronic requirements of the emitter (contact layers, minimizing current density and Ohmic heating, etc.) often conspire to significantly limit the possible enhancement, in addition to adding significant parasitic effects detrimental to device performance \cite{Khurgin:14}. In the device described here, however, the monolithic nature of the device architecture, combined with the longer wavelength operation, allow for placement of the emitters in the near field of the plasmonic material without compromising device operation.

Two LED structures, each containing five layers of In(Ga)Sb quantum emitters embedded in an InAs PIN-junction, were grown by molecular beam epitaxy on doped n-type InAs ‘virtual substrates’: the first, a control sample, was grown on a moderately-doped n-InAs backplane (n-doped ${\sim}$2 x 10$^{18}$ cm$^{-3}$), and the second, our n${^{++}}$ device that monolithically integrates electrically pumped material and long wavelength metal, on heavily doped n${^{++}}$ InAs (active concentration ${\sim}$1.2 x 10$^{20}$ cm$^{-3}$).  In both cases, the top of the virtual substrate lies 100 nm below the first In(Ga)Sb emitter layer; band diagrams and layer structures of the devices, as well as bright-field transmission electron microscope (TEM) images of the quantum emitters, are shown in Fig. 1. Each active region was capped with a lattice-matched p-AlAs${_{0.16}}$Sb${_{0.84}}$ barrier, which has a significant conduction band offset with InAs, confining electrons injected into the active region and mitigating parasitic surface recombination \cite{Schubert}. To promote current spreading and decrease contact resistance the barriers were capped with a p${^{++}}$ InAs layer (doped ${\sim}$1 x 10$^{19}$ cm$^{-3}$). \textcolor{black}{Both samples were processed using traditional UV photolithography, wet-etch, metallization, and lift-off techniques to form LED mesas (700 ${\mu}$m x 600 ${\mu}$m) with a window top contact of Ti/Pt/Au (50 nm/60 nm/140 nm) via e-beam evaporation.  Backside contacts of Ti/Au (10 nm/ 200 nm) were deposited on the device substrate. LED samples were indium-bonded to copper blocks, wire-bonded to Au-coated ceramic stand-offs, and then mounted in a temperature-controlled pour-fill cryostat.}

\begin{figure}[t]
\centering
\includegraphics[width=\linewidth]{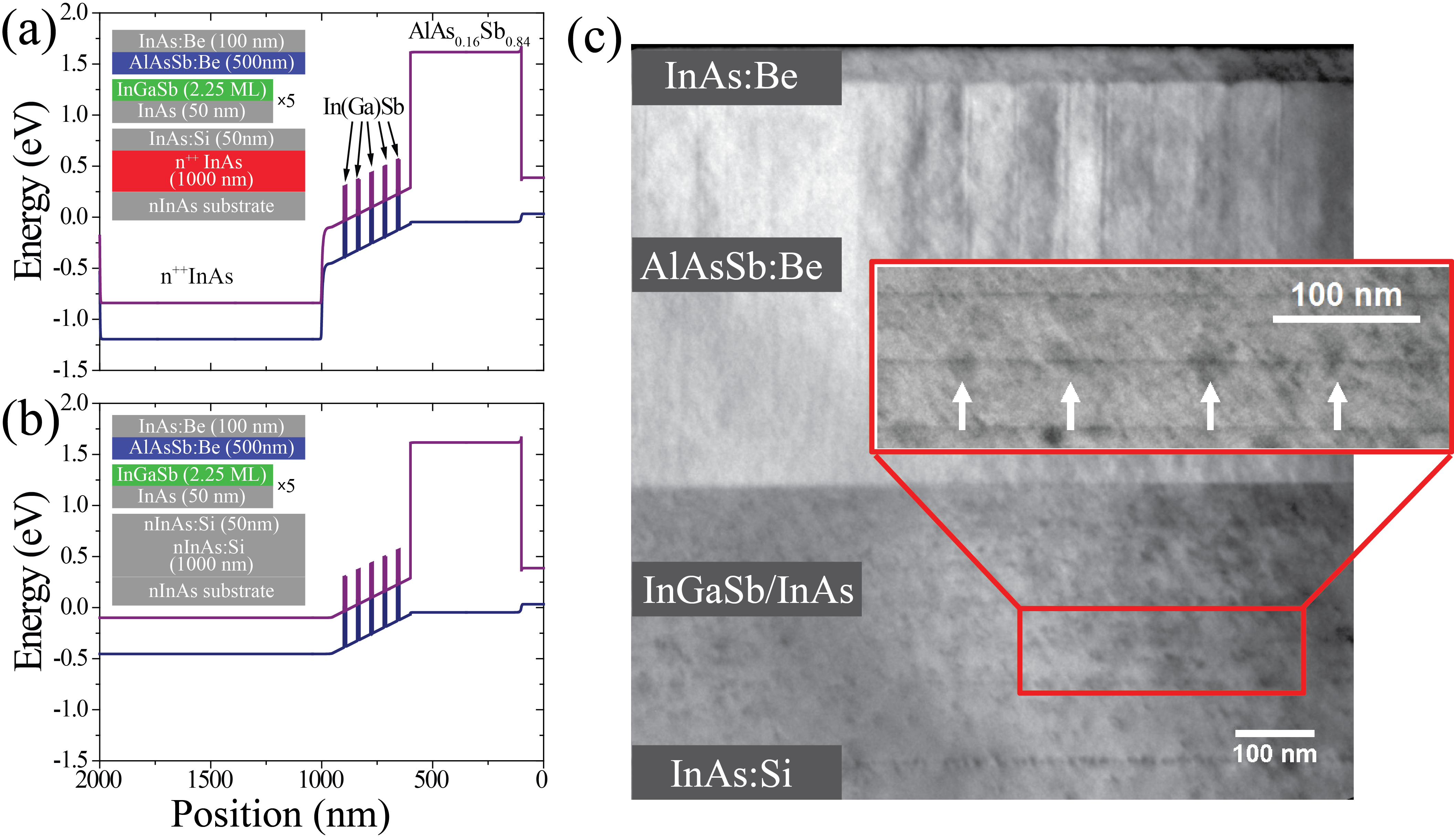}
\caption{Band diagrams for unbiased (a) cavity-enhanced In(Ga)Sb LED and (b) control In(Ga)Sb LED. \textcolor{black}{Insets: layer structure of each device. (c) Representative bright-field TEM of the cavity-enhanced In(Ga)Sb LED. Arrows indicate strain fields from discrete quantum dots.}}
\label{Figure1}
\end{figure}

\begin{figure}[t]
\centering
\includegraphics[scale=.65]{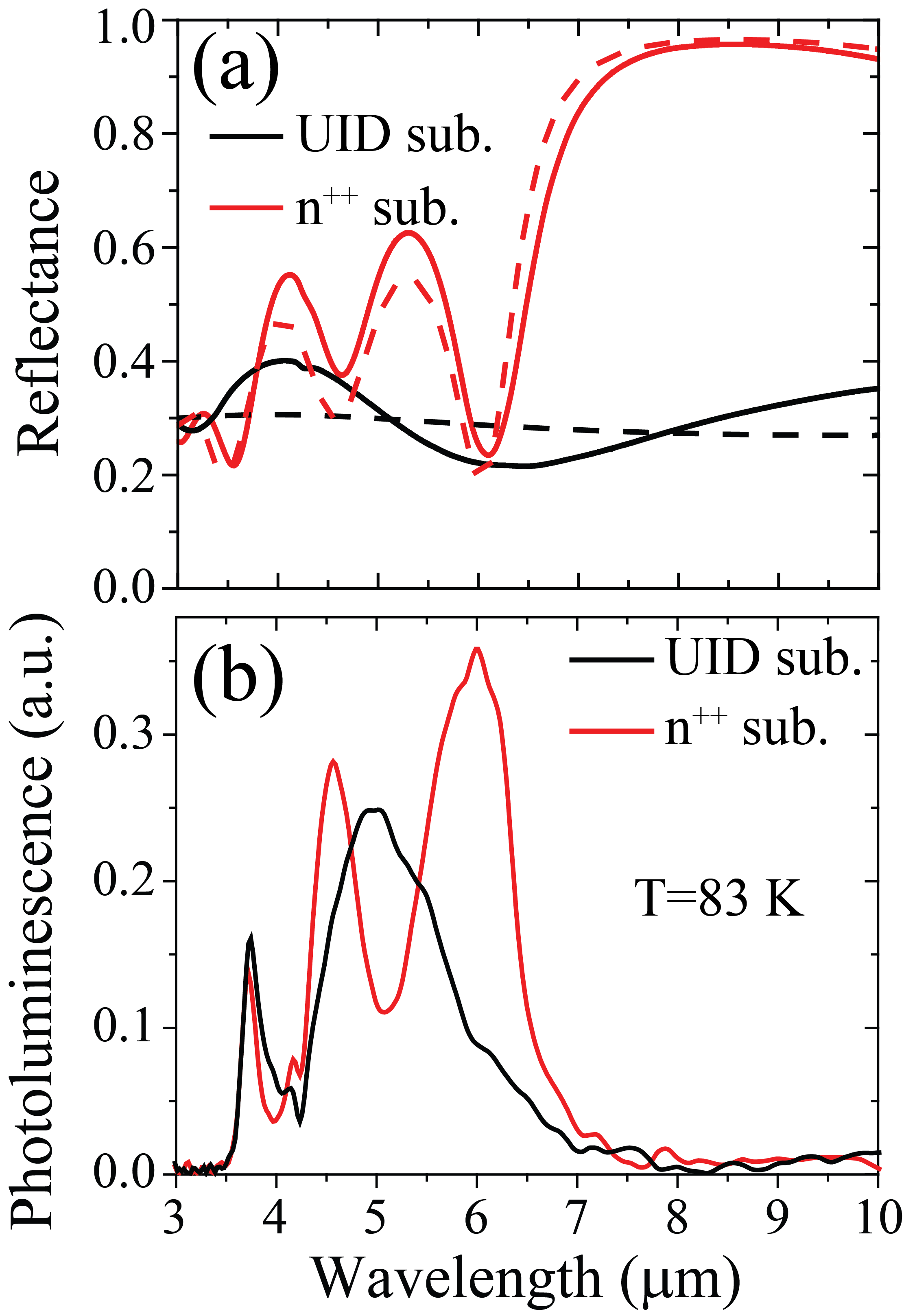}
\caption{(a) Room temperature experimental (solid) and modeled (dashed) reflectance of In(Ga)Sb quantum emitter samples grown on highly-doped (red) and moderately-doped control (black) InAs virtual substrates. (b) Low temperature (83 K) photoluminescence measured from the InGaSb diode samples with control (black) and highly-doped (red) substrates.}
\label{Figure2}
\end{figure}

The reflectance spectra of both devices were measured [Fig. 2(a)] and fitted using a transfer-matrix method, treating the doped semiconductor layer as a Drude plasmonic material of permittivity
\begin{center}
$${\epsilon_o(\omega)=\epsilon_\infty \left( 1 - \frac{\omega_p^2}{\omega^2 + i\gamma \omega} \right)}$$
\end{center}
with the plasma frequency (${\omega_p = 2 \pi c / \lambda_p}$) and scattering rate (${\gamma}$) of the virtual substrate as the fitting parameters, from which we extracted the plasma wavelength and scattering rate of the n${^{++}}$ virtual substrate of the n${^{++}}$ monolithic device (${\lambda_p}$ = 6.25 ${\mu}$m, ${\gamma}$ = 5 x 10$^{12}$ Hz). Qualitatively, the heavily doped sample showed the spectral features expected of a three-layer air-dielectric-Drude metal system: the strong reflectance at long wavelengths and the noticeable reflection dip at 6 ${\mu}$m corresponding to a leaky ${\lambda}/4n$ cavity mode set up by the three-layer system \cite{Streyer:13}.

\begin{figure}[t]
\centering
\includegraphics[scale=.08]{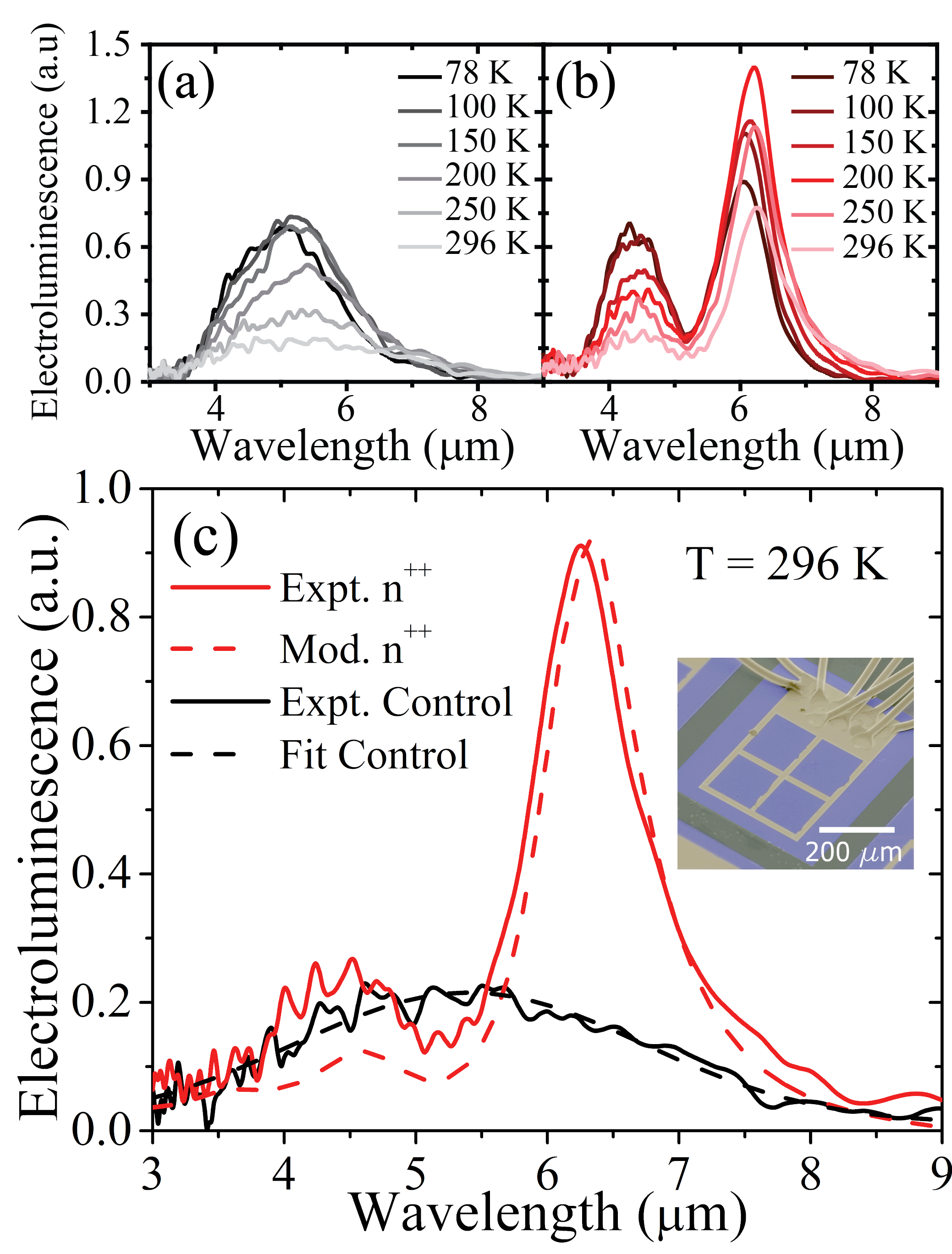}
\caption{Temperature-dependent EL spectra of the (a) five layer control device and the (b) cavity-enhanced device with the n${^{++}}$InAs backplane (c) Modeled (dashed, assuming ${q_i = 5 \%}$) and experimental room temperature comparison of n${^{++}}$ (black) and control (red) devices showing strong cavity enhancement around 6 ${\mu}$m. Inset: False colored SEM micrograph of a fabricated cavity-enhanced In(Ga)Sb LED, the LED mesa is in blue and the Ti/Au contact pads and Au wire bonds are in yellow.}
\label{Figure4}
\end{figure}

\textcolor{black}{Both samples were optically characterized by Fourier transform infrared (FTIR) amplitude modulation step scan photoluminescence (PL) spectroscopy, using a 980 nm pump laser modulated at 10 kHz.  The resulting detector signal was demodulated by a lock-in amplifier synchronized to the pump source modulation frequency.} The low temperature PL spectrum of the control device [Fig. 2(b)] is dominated by the broad continuous QD emission spectrum centered at ${\lambda}$ = 5 ${\mu}$m, with PL spectrum cut off at ${\lambda}$ = 3.6 ${\mu}$m by a long pass filter, used to block the InAs bandedge emission and reflected pump laser light. The PL from the highly-doped (n${^{++}}$) device showed a markedly different spectrum [Fig. 2(b)], with a clear enhancements (suppressions) of the control PL spectrum commensurate with the reflection minima (maxima) observed for the n${^{++}}$ device (which, again, come from the leaky cavity formed by the air/dielectric/Drude metal system).  The most prominent enhancement is observed at the long wavelength tail of the QD emission, where the modulation of the emission from the n${^{++}}$ sample included a strong enhancement of emission, by a factor of approximately 3.9 ${\times}$ in photoluminescence.

To elucidate this effect, we modeled our system using a Dyadic Green’s function formalism, incorporated into a transfer matrix method solver \cite{Novotny,Brueck:00,Nordin:20}, positioning our emitter at the five QD layer positions. \textcolor{black}{We used an internal quantum efficiency for the QDs (${q_i}$) as a fitting parameter in our model, with ${q_i = 5 \%}$, in line with typical emission efficiencies of MWIR type-II emitters \cite{Biefeld:1998,Muhowski:19}. Further discussion of our model and the data fitting approach can be found in the supplemental information.}

\begin{figure}[t]
\centering
\includegraphics[scale=.12]{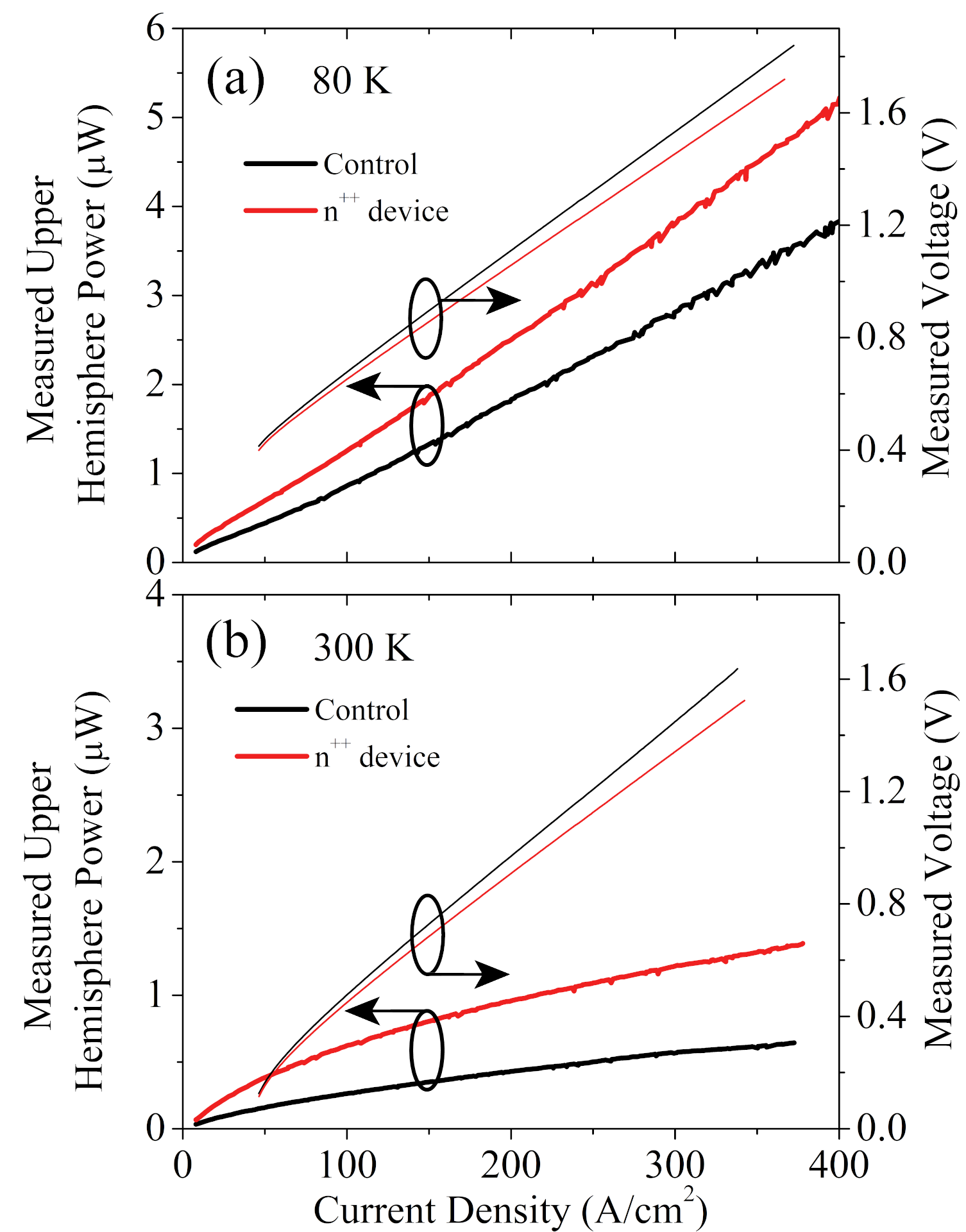}
\caption{\textcolor{black}{Measured} L-I-V Characteristics of the control (black) and n${^{++}}$ (red) devices at (a) low temperature (80 K) and (b) room temperature (300 K). }
\label{Figure5}
\end{figure}


\textcolor{black}{Electroluminescence (EL) was also characterized using a FTIR amplitude modulation step scan with the sample modulation driven by an Agilent 8114a pulsed current source. For temperature-dependent EL measurements, the samples were pulsed at 400 mA in “quasi-CW” mode (a duty cycle of 50$\%$) with a repetition rate of 10 kHz. For power measurements, light from the devices, modulated at 1 kHz and 50$\%$ duty cycle, was collected by a $f/2$ ZnSe lens and then focused onto a calibrated HgCdTe detector using a second $f/2$ ZnSe lens. The effective responsivity of the detector was scaled by the transmission through the ZnSe optics and integrated over the corresponding angle for $f/2$ collection to compute the upper hemisphere power \cite{Vincent}. Both ZnSe lenses were 1-inch diameter and had a 2-inch focal length. A lock-in amplifier was used to demodulate the collected HgCdTe signal.} Temperature-dependent electroluminescence (EL) of the control sample [Fig. 3(a)] showed a significant decrease in emitter efficiency with increasing temperature, expected for narrow bandgap emitters \cite{Jung:17}. The n${^{++}}$ device, however, showed remarkably different temperature dependence, as shown in Fig. 3(b). While the short wavelength feature of the n${^{++}}$ device (centered around 4.5 ${\mu}$m) decayed in intensity in a similar manner to the control sample’s emission, with intensity decreasing 3 ${\times}$ from 78 K to 298 K, the long wavelength emission peak from the n${^{++}}$ device showed only a ${13\%}$ decrease in emission intensity between the low temperature and room temperature measurements, which is highly unusual for mid-IR emitters. Applying our enhancement model to the control sample emission (fitted to a Gaussian lineshape) and assuming ${q_i = 5 \%}$, we can accurately predict the room temperature EL spectrum of the n${^{++}}$ device, shown in Fig. 3(c). The electroluminescence showed a strong enhancement of emission at peak wavelength, increasing by a factor of approximately 5.6 ${\times}$. The calculations that predict the observed EL enhancement of the emission are quite accurate when compared to experimental results.

Light-current-voltage (LIV) measurements of the two emitters (Fig. 4) demonstrates a near uniform enhancement of emission across all pump currents measured at low temperature (80 K) and room temperature (300 K). Peak upper hemisphere emitted power at low temperature, shown in Fig. 4(a), exceeds 5 ${\mu}$W. Room temperature peak upper hemisphere emitted power for the n${^{++}}$ device, shown in Fig. 4(b), was measured to be 1.45 ${\mu}$W, which is already comparable to state-of-the-art commercial mid-IR LEDs at 6 ${\mu}$m \cite{IoffeeLED}.

We report the first all-epitaxial integration of a plasmonic “designer metal” and emitter to produce an electrically-pumped, cavity-enhanced LED.  The use of a designer metal virtual substrate resulted in a significant spectral modulation of emission intensity, little to no decrease in emission intensity over a range of 200 K, and a strong enhancement of emission, which we modeled and determined to result from a combination of Purcell enhancement and the leaky cavity formed by our three layer system. The plasmonic virtual substrate device showed upper hemisphere powers of 1.45 ${\mu}$W at 300 K, comparable to state-of-the-art mid-IR LEDs \cite{IoffeeLED}. The device architecture presented here offers a new approach to enhance mid-IR optoelectronic devices, taking advantage of our ability to engineer both our designer metals and our quantum emitters in the same epitaxial material platform. In addition to the potential practical applications for mid-IR optical systems, the devices introduced here offers unprecedented opportunities for exploring and engineering light-matter interactions and new device architectures leveraging monolithically-integrated designer plasmonic materials. 

\section*{Funding Information}
The authors gratefully acknowledge support from the National Science Foundation (Grant Nos. ECCS-1926187, DMR-1508783, DMR-1629330, and DMR-1508603) and the Defense Advanced Research Projects Agency through the NLM program.

\section*{Disclosures}
\noindent\textbf{Disclosures.} The authors declare no conflicts of interest.
\bigskip

\noindent See supplementary materials for supporting content.


\bibliography{sample}

\bibliographyfullrefs{sample}


\end{document}